\documentclass{camera}
\usepackage{epsfig}
\usepackage{amssymb,natbib}

\def\annihilateur{1 \mbox{E 1740.7-2942}}

\def\grs{\mbox{GRS 1915+105}}

\def\xtejodh{\mbox{XTE J1118+480}}
\def\xtejqcq{\mbox{XTE J1550-564}}

\def\Msol{\mbox{ }M_{\odot}}

\def\deg{^{\circ}}

\def\asec{^{\prime \prime}}

\def\rg{r_{\scriptsize g}}
\def\ltsima{\; \buildrel < \over \sim \;}
\def\simlt{\lower.5ex\hbox{\ltsima}}            % < over MMM
\def\gtsima{\; \buildrel > \over \sim \;}
\def\simgt{\lower.5ex\hbox{\gtsima}}            % > over MMM

\begin{document}

%%%%%%%%%%%%%%%%%%%%%%%%%%%%%%%%%%%%%%%%%%%%%%%%%%%%%%%%
% The title, only the first letter capitalized; if you want to split it in
% two or more lines, put a \\ macro at each line break
% example: 
%   \title{Title: first line\\ second line}
%
\title{The role of microquasars in astroparticle physics}

%%%%%%%%%%%%%%%%%%%%%%%%%%%%%%%%%%%%%%%%%%%%%%%%%%%%%%%%
% The author(s), separated by commas; do not put a
% comma before the last author, use instead the \and
% macro which produces a normal ``and'' in the
% caps/small caps context
%
\author{S. Chaty}

%%%%%%%%%%%%%%%%%%%%%%%%%%%%%%%%%%%%%%%%%%%%%%%%%%%%%%%%
%
\organization{AIM - Astrophysique Interactions Multi-\'echelles
(Unit\'e Mixte de Recherche 7158 CEA/CNRS/Universit\'e Paris 7 
Denis Diderot),
CEA Saclay, DSM/DAPNIA/Service d'Astrophysique, B\^at. 709,
L'Orme des Merisiers, FR-91 191 Gif-sur-Yvette Cedex, France}

\maketitle

\begin{abstract}
I present an overview of past, present and future research on
microquasars and jets, showing that microquasars, i.e. galactic jet
sources, are among the best laboratories for high energy phenomena and
astroparticle physics.  After reminding the analogy with quasars, I
focus on one of the best microquasar representatives, probably the
archetype, namely GRS 1915+105, and present accretion and ejection
phenomena, showing that only a multi-wavelength approach allows a
better understanding of phenomena occurring in these
sources. Thereafter, I review jets at different scales: compact jets,
large-scale jets, and the interactions between ejection and the
surrounding medium. I finish this review by showing
that microquasars are good candidates to be emitters of astroparticles: 
very high energy photons, cosmic rays and neutrinos. 
\end{abstract}

%%%%%%%%%%%%%%%%%%%%%%%%%%%%%%%%%%%%%%%%%%%%%%%%%%%%%%%%
% Write the text starting from here and using the usual
% LaTeX commands.
%

\section{The pre-microquasar era: SS 433} \label{prelude}

In 1979 was discovered the microquasar prototype: SS 433, a
high-energy source exhibiting precessing jets at frame velocity $0.26
c$, with emission lines observed in the optical, showing that the jet
content was baryonic \citep{margon:1984}. SS 433 is surrounded by a
supernova remnant: W50, and there are clear signs of interaction
between SS 433 jets and W50 nebula (see e.g. \citeauthor{dubner:1998}
\citeyear{dubner:1998}).  The question which arose was then: how can a
galactic object eject matter at such relativistic velocities
($\Gamma$=1.04)? This object exhibited such unusual
properties, that it was probably impossible to foresee that, two
decades later, jet sources would become quite common. SS 433 had
everything of a microquasar, apart from the name.

\section{The microquasar era: analogy with quasars} \label{youth}

In 1990, the {\it SIGMA} telescope, orbiting on board {\it Granat}, was
launched. It was designed to observe galactic black hole candidates,
because its observing energy band corresponded to the energy released
by accretion around compact objects.  In 1992 the first so-called
microquasar, $\annihilateur$, was identified
\citep{mirabel:1992b}. This source was exhibiting bipolar radio jets
spread over several light-years. This was the first such
observation in our Galaxy, however jets had been already observed emanating
from distant galaxies. Therefore this observation made clear the
existence of a morphological analogy between quasars and microquasars.

\begin{figure}
\centerline{\psfig{file=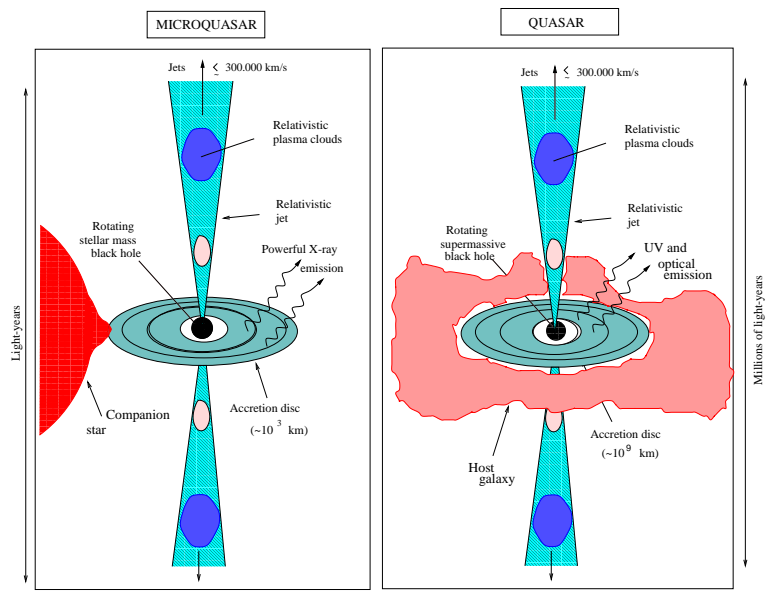,angle=-90.,width=10.cm}}
% quasar_microquasar_english.ps
\caption[]{\label{quasar_microquasar} Schematic view illustrating 
analogies between quasars and microquasars.  Note
the different mass and length scales between both types of objects 
\citep{chaty:1998}.  }
\end{figure}

Although there is no clear definition of a microquasar, we can
characterise it as a galactic binary system --constituted of a compact
object (stellar mass black hole or neutron star) surrounded by an
accretion disc and a companion star-- emitting at high-energy and
exhibiting relativistic jets. A schematic view of a
microquasar, compared with quasars, is given in Figure
\ref{quasar_microquasar}. Taking this broad definition, we observed
nearly $20$ microquasars in our Galaxy, and it is one of the main
subjects of study by current space missions. Since each component
 of the system emits at different wavelengths, it is
necessary to undertake multi-wavelength observations in order to
understand phenomena taking place in these objects.

In 1992 the WATCH/GRANAT telescope discovered the black hole candidate
GRS 1915+105 \citep{castro:1994}, which would become the archetype of
microquasars.  Two years later, by observing this source with the VLA
(arcsec scale), \cite{mirabel:1994a} detected apparent superluminal
motions, while frame velocity was $v\sim0.92c$.  It became then
rapidly clear that the advantages of microquasars compared to quasars
were that i) they are closer, ii) it is possible to observe both
(approaching and receding) jets, and iii) the accretion/ejection timescale is
much shorter.  After this observation of superluminal motions, the
morphological analogy with quasars became stronger, and the question
was then: is this morphological analogy really subtended by physics? 
If the answer is yes, then microquasars really are
``micro''-quasars. For instance, there should exist microblazars
(microquasar whose jet points towards the observer), in order to
complete the analogy with quasars.

We will see in the following that this quasar/microquasar analogy
became rapidly very fruitful, the field of quasars benefiting of
microquasars, and vice versa. For instance, because accretion/ejection timescale is proportional to black hole mass, it is easier
(because faster) to observe accretion/ejection cycles in microquasars
than in quasars \footnote{Characteristic timescale of phenomena
occurring very close to the last stable orbit around the black hole of
mass $M$ is given by $\tau \sim \frac{\rg}{c} \sim M$, where $\rg$ is
the Schwarzschild radius.  Therefore, this timescale is proportional
to the mass of the black hole. If a stellar mass black hole exhibits
accretion/ejection cycles of a few minutes, a supermassive black hole
will exhibit corresponding cycles on a few thousands of years.}. On
the other hand the understanding of ejection phenomena in microquasars
have largely benefited from jet models developed for active
galaxies.

%---------------------------------------------------------------------
%---------------------------------------------------------------------

\section{The golden age of microquasars: accretion and ejection}

\begin{figure}
\centerline{\psfig{file=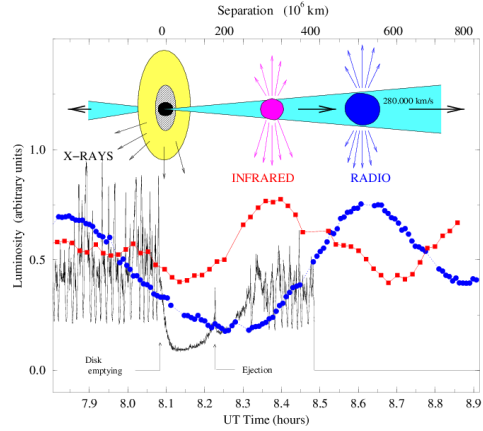,angle=-90.,width=8.cm}}
% scenario_english2.ps
\caption[]{\label{grs_sept_french} Observation of the link between 
accretion and ejection. X-ray, NIR and radio lightcurves of $\grs$
during the 1997 September 9 multi-wavelength observation campaign. 
The disappearance of the internal part of the accretion disc (decrease in
the X-ray flux) is followed by an ejection of relativistic plasma
clouds (oscillation in the NIR and radio) 
(\citeauthor{chaty:1998} \citeyear{chaty:1998}; 
\citeauthor{mirabel:1998a} \citeyear{mirabel:1998a}).}
\end{figure}

\begin{figure}
\centerline{\psfig{file=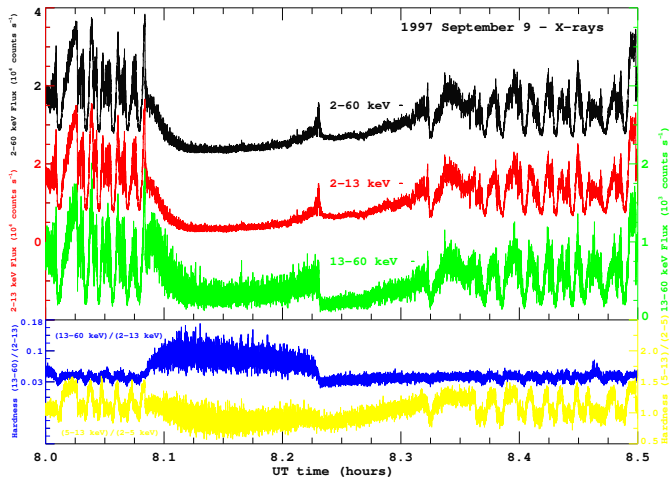,angle=90.,width=9.cm}}
% xte_9sept1997_zoom_tout-english.ps
\caption[]{\label{xte_9sept1997_zoom} Same observations as above, with only
X-ray observations, and enlarged on the UT interval
[8.0-8.5] hours. From top to bottom: 2-60 keV, 2-13 keV
and 13-60 keV X-ray flux; hardness ratio $\frac{13 - 60 keV}{2 - 13
keV}$ and $\frac{5 - 13 keV} {2 - 5 keV}$. These observations
suggest that it is a part of the corona which is ejected at the time
of the X-ray spike \citep{chaty:1998}.}
\end{figure}

GRS 1915+105 will once again play an important role in the
understanding of microquasars.  In 1997, after performing
multi-wavelength observation campaigns of this source, the link
between accretion and ejection was discovered (\citeauthor{chaty:1998}
\citeyear{chaty:1998}; \citeauthor{mirabel:1998a}
\citeyear{mirabel:1998a}). Examining Figure \ref{grs_sept_french}, we
can see the disappearance of the internal part of the accretion disc,
shown by a decrease in the X-ray flux, followed by an ejection of
relativistic plasma clouds, corresponding to an oscillation in the
near-infrared (NIR) and then in the radio, the cloud becoming
progressively optically thin.  The analysis of X-ray fluxes and
hardness ratios, shown in Figure \ref{xte_9sept1997_zoom}, suggests
that it is mainly the part emitting at higher energy which is ejected
at the time of the X-ray spike. This supports the interpretation that
part of the corona (surrounding the compact object in the central part
of the accretion disc) is ejected during this cycle
\citep{chaty:1998}. Each of these accretion/ejection cycles last for
$\sim10$ min, and they are recurrent, occurring every $\sim 30$--$45$
min.  Not only it is interesting to point out that these observations
had not been performed on quasars, even after nearly 40 years of
study, but also that for the first time microquasars were taking over
on the quasars, bringing new discoveries.  Five years later, similar
phenomena would be reported on the quasar 3C120, compiling 3 years of
observations \citep{marscher:2002}. These observations from both types
of objects confirmed that the morphological quasar/microquasar analogy
was subtended by physics \footnote{Another compelling evidence of this
analogy is given by the supermassive black hole at the centre of our
Galaxy: with a mass of $3.6\times 10^{6}\Msol$ it exhibits a few tens
of minutes NIR quasi-periodic oscillations (QPOs;
\citeauthor{genzel:2003} \citeyear{genzel:2003}), when stellar mass
black holes exhibit a few millisecond X-ray QPOs, consistent with the
mass ratio.}.

%-----------------------------grs_sept_total_french--------------------
%-----------------------------------------------------------------------

%--------------------------------------------------------------------
%---------------------------------------------------------------------------

%\section{The golden age of microquasars: emission processes}

We will not discuss here the different accretion and ejection
models, but refer the reader to e.g. \cite{fender:2001a} for a
description of these models and how they relate to different ejection
states. We simply remind that the standard model is
constituted of thermal emission coming from a multicolour black body accretion
disc and of non-thermal emission of plasma corona, and that jets are
observed during low/hard states (historically referring to X-rays). 
%Some microquasars emit at high energy, their emission being
%dominated by a power law (spectral index 2.5-3), without any cut-off
%\cite{grove:1998}. 
Concurrent models invoke jet synchrotron emission from radio to
X-rays.  Therefore the main uncertainty in this domain concerns the
underlying physical process: comptonization or synchrotron? An answer
might be given by polarisation observations.  High energy instruments
do not allow this yet, and NIR polarimetric observations are still
beginning.  \cite{dubus:2006b} report NIR polarimetric observations of
the microquasar $\xtejqcq$, performed in 2003 at ESO/NTT.  These
observations were performed on the decline (at $\sim2.5$ count/s) of a
small amplitude outburst peak (4.5 count/s) detected by {\it
Rossi-XTE}/ASM \citep{sturner:2005} and which lasted about a month: it
was 3.2 mag brighter in NIR than in quiescence. $\xtejqcq$
polarisation is inconsistent with other stars of the field of view at
the $2.5 \sigma$ level, suggesting an intrinsic NIR polarisation
p=0.9--2.0\% perhaps due to synchrotron emission from the jet,
associated with the outburst \citep{dubus:2006b}.

\begin{figure}
%\setlength{\unitlength}{1.0cm}
%\begin{picture}(7.,7.)(-2.,-7.5)
%\special{psfile=sed_tout+mod_fig.ps hscale=40 vscale=40 angle=-90}
%\end{picture}
\centerline{\psfig{file=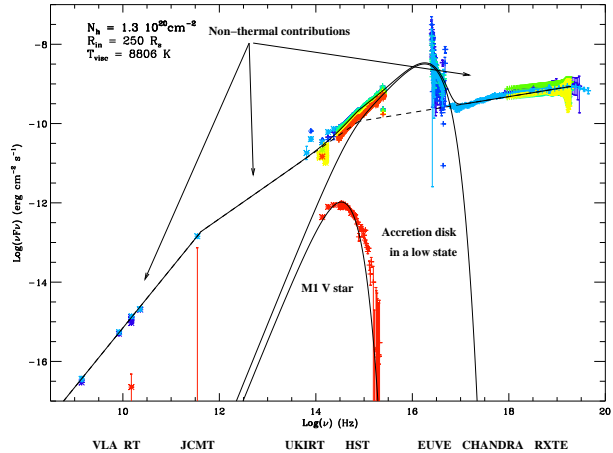,angle=-90.,width=10.cm}}
% sed_tout+mod_fig.ps
\caption[]{\label{sed_1118} Spectral energy distribution of the
microquasar $\xtejodh$ which can be described by the sum of emission
from the multicolour blackbody accretion disk, a non-thermal
contribution from an outflow and the M1 V companion star \citep{chaty:2003b}.}
\end{figure}

To understand accretion/ejection models, it is therefore necessary to
undertake a multiwavelength approach and get the spectral energy
distribution (SED) of various sources.  There is a small number of
microquasars for which this has been done intensively, the jet source
and black hole XTE J1118+480 being one of the best examples, favoured by the
very low absorption on its line of sight \citep{chaty:2003b}. In Figure
\ref{sed_1118} I report the SED of this
source, including 6 different epochs of simultaneous multi-wavelength
observations from radio to X-rays, performed with 8
different instruments. On this Figure I overplot the thermal
emission of the multicolour black body accretion disc, the 
emission from the companion star, and non-thermal emission which
appears to be necessary to account for radio, NIR and X-ray
domains.  In \cite{chaty:2003b} it has been shown, by using a
non-linear Monte-Carlo simulation, that the presence of hot spherical
plasma in the centre can account for the emission of the source from
optical to X-rays.  However other models show that this emission can
also be described by a jet emitting from radio to X-rays, as in the
case of active galaxies \citep{markoff:2001}.  This question about the
jet contribution is therefore still a matter in the debate.
%, and we will need more sensitivity and temporal resolution to eventually solve it.

It is interesting to compare XTE J1118+480 and GRS 1915+105
SEDs. During large multiwavelength campaigns from radio to hard
X-rays, \cite{ueda:2002} and
\cite{fuchs:2003} have shown the presence of a flat radio spectrum,
during the ``plateau'' (or low/hard) state of GRS 1915+105.  They
also confirm that the jet contributes to the emission in the
NIR domain. A comparison of the accretion/outflow energy ratio of
both sources XTE J1118+480 \& GRS 1915+105 shows that they both fall
into the regime of radio-quiet quasars \citep{chaty:2003b}.

Simultaneous multi-wavelength observations of both types of objects,
namely microquasars and quasars, will eventually bring severe
constraints on accretion-ejection models (e.g. Blandford-Payne,
Blandford-Znajek, Magneto-Rotational Instability...), and on the
nature of the jets (are they baryonic or leptonic?). For instance,
putting together radio and X-ray observations suggests that a coupling
exists between both domains, $F_{\rm rad} \propto F_{\rm X}^{+0.7}$, for
galactic \citep{gallo:2003} and extragalactic jet sources
\citep{falcke:2004}, but a
good understanding of this coupling still misses.
%This shows that there is no
%strong Doppler amplification: low velocity jet ($\beta < 0.8c, \Gamma < 2$).
%  Is there a coupling at higher energy?
Some answers might also come from the detection of (Doppler- shifted?) 
annihilation emission lines, and also from observations of QPOs in
microquasars.

%------------------------------XTE J1118--------------------------
%---------------------------------------------------------------------------

\section{The hidden face of microquasars: interaction with their surroundings}

Jets of microquasars can be observed at different scales,
corresponding to different sizes and energy outputs involved.
Observations of sporadic ejection at large scale were performed
first, as described in Section \ref{youth}.  A steady compact jet
has been observed in a few microquasars, for instance in GRS 1915+105
(at the milli-arcsec scale, where 10mas = 1AU;
\citeauthor{dhawan:2000b} \citeyear{dhawan:2000b};
\citeauthor{fuchs:2003} \citeyear{fuchs:2003};
\citeauthor {ribo:2004} \citeyear{ribo:2004}).
Since these jet sources eject a large amount of matter in the
interstellar space, which is far from being empty, it appeared fruitful
to look for interactions between jets and surroundings of the
microquasar.  The first example is $\annihilateur$, which exhibits a
steady jet, probably due to the braking of its continuous jet in the
interstellar medium.  The signature of such an interaction might be
the observation, directly in the jets, of a narrow annihilation line
at 511 keV, due to $e^+$ colliding with the interstellar medium
\footnote{Annihilation lines have been reported on this source
(therefore also called ``the great annihilator of the Galaxy'') but
likely coming from the central source, and therefore related
to the accretion process \citep{bouchet:1991}.}.
Large-scale jets are now regularly observed in X-rays. \cite{corbel:2002} have
observed such jets emanating from the microquasar $\xtejqcq$, at 
$45\asec$ of the central source. To emit at such energy,
the particles have to be accelerated up to TeV energy, again
strengthening the analogy with quasars.  
%However, the question of the e- maximal energy is still open.

By studying the interactions between the jets and the interstellar medium,
one can not forget GRS 1915+105: always active, transient, and the
place of very energetic ejection.  Such interactions in the
surroundings of GRS 1915+105 had already been suggested nearly 10
years ago by \cite{mirabel:1996a}.  In August 1995, during a strong
and long X-ray outburst of GRS 1915+105, the radio source was resolved
in 2 jets, and the NIR emission increased significantly between 2 and
5 days after the radio burst. \cite{mirabel:1996a} interpreted this as
the presence of an extended cocoon of dust, heated by ejection.  This
dust in the surroundings of this microquasar was later confirmed by
{\it Chandra} \citep{lee:2002} and {\it ISO} \citep{fuchs:2001}
observations.  However it is still unknown if the cocoon has been
created by previous ejection, or by accumulation of ISM dust.  What
about the surroundings of GRS 1915+105, at larger scale?  A
low-resolution centimetre map exhibits two sources aligned with the
central source \citep{chaty:2001a}. By observing them at higher
resolution, \cite{chaty:2001a} discovered a strange non-thermal feature in the
south-east lobe, which might be a synchrotron signature of
interactions between jets and ISM. However, they
concluded that even if, based on the energy output, the interaction is
a possibility, there is no observational fact allowing to confirm that
this strange feature is the signature of interaction between jets and
interstellar medium.

Finally, all these observations of jets bring us to another important
question in the field of jet sources: are the jets a propagation
of plasma clouds or of shock waves?  The first
interpretation is usual among the microquasar community, and the
second one among the extragalactic community. By applying 3C273 model
to GRS 1915+105, \cite{turler:2004} have shown that ejection in GRS
1915+105 could be described as the propagation of a shock wave forming
at 1AU, with dissipative stream at $v=0.6c$.

\section{Microquasars and their role in astroparticle physics}

In this review I have shown that microquasars are the site of
accretion, ejection and interaction of jets with the interstellar
medium. Microquasars are therefore high-energy laboratories which
encompass all necessary ingredients in order to emit
astroparticles: VHE photons, cosmic rays and neutrinos.

\subsection{Microquasars as emitters of VHE photons}

Many models exist which predict the emission of VHE photons by
the microquasars. A recent one is the broad-band leptonic model 
for gamma-ray emitting microquasars by \citet{bosch-ramon:2006},
in which the jet is dominated dynamically by cold protons, radiatively
by relativistic leptons, and the magnetic field is at equipartition. 
In this model the emission from radio to VHE is due to 
synchrotron, relativistic Bremsstrahlung and Inverse Compton processes. 
It provides predictions about the shape of the SEDs and points 
to microquasars as VHE sources.

There are some observations of GeV emission from EGRET (Paredes et
al. 2000) to VHE TeV $\gamma$-rays with HESS (Aharonian et al. 2005)
emanating from the microquasar candidate LS 5039. A second example is the detection of variable
VHE $\gamma-ray$ emission from the microquasar candidate LS I +61 303, with
MAGIC. However, it is not clear if the high energy emission is coming
from the jets of these sources, or from the interaction between
radiation of the compact object (if it is a neutron star)
 and the wind of their early spectral type
companion star as proposed by \cite{dubus:2006a}\footnote{This kind of object has been nicely called ``pulsars in
disguise'' by this author.}.  
In the latter case no jet would be
needed, and these sources would not be microquasars.

\subsection{Microquasars as neutrinos emitters}

The model by \citet{romero:2005} predicts neutrino and gamma-ray
emission from misaligned microquasars, if the jet is misaligned
with the perpendicular to the orbital plane (which could be usual: 
for instance, the jet is 
inclined of $35 \deg$ from orbital plane for V4641 Sgr; there is some 
precession in SS 433, LSI +61 303, and Cygnus X-3), and if the donor
star is an early-type star. If both conditions are met,
then the jet collides with the stellar wind, and a
 standing shock between the compact object and the stellar
surface is created. Finally, if the jet has an hadronic content, TeV protons 
might diffuse into an inner, dense wind leading to $\gamma$-ray
emission.
This model predicts an enhancement and periodic variability of $\gamma$-ray TeV signal (which could be detected by HESS, MAGIC, Veritas) and also
 a neutrino signal at 3 $\sigma$ (ICECUBE, AMANDA, ANTARES) 
for an observation time of 15 years of a source located at 2 kpc... only if
there is a close alignment of the jet with the line of sight, 
in order to obtain a duty cycle of 20\%.

Therefore even if the microquasars are emitters of TeV neutrinos, 
their possible detection will probably only be feasible with km$^2$ 
neutrino telescopes. To see what could be the signal, we can take 
the model of \citet{distefano:2002}, where there is photopion production 
in the jet, with the conditions that the jets are protonic, 
and that a fraction of a few percents of the jet energy is dissipated 
on a sufficiently small scale.
In this case, we would expect a signal of 252 neutrinos coming from SS 433,
 with a 1 year integration time.
We could probably even identify new microquasars by their 
neutrino and $\gamma$-ray emission, if they exhibit 
large bulk Lorentz factors, and if the jets are directed along our 
line of sight.
 
\subsection{Microquasars as cosmic rays emitters}

If the jets contain cold protons and heavy ions (as it is the case in
SS 433), cosmic rays from microquasars could represent a narrow
component at 3-10 GeV to the Cosmic ray spectrum \citep{heinz:2002}.
For instance, a single microquasar as GRS 1915+105, if it is active 
for $10^7$ years, with a luminosity of $10^{38}$ergs/s, and located
at a distance of 1 kpc, would produce detectable signal in the galactic
CR background spectrum.
On the other hand, their prediction is that if there is no such
detection, this means that the jets are leptonic...

\section{Conclusions}

We have seen that microquasars are excellent laboratories for high energy
 physics, and are the key to many still pending questions, related to
accretion-ejection mechanisms, interaction between the jets and interstellar
medium, and their propagation.
To answer to these questions, it will be necessary to simultaneously
study multi-wavelength emissions from various microquasars at
different stages. This will also allow to better measure
what is the contribution in the SED of non-thermal (synchrotron) 
emission coming from the jets. 
One of the key questions, perhaps the most important one, is whether 
the jets are baryonic or leptonic. Indeed, we have seen
that microquasars will not have the same role in astroparticle physics
in both cases, since there is much more chance for them to be VHE photons,
cosmic rays and neutrinos emitters if the jets are of baryonic nature.

We are at an era where the whole electromagnetic spectrum can be
explored, from the radio to VHE photons, along with cosmic rays,
and we will be able soon to
observe neutrinos. Putting everything together and linking 
these observations with theory and models might help us 
in a better understanding of these exciting sources. 

%But obviously we will also need some more Vulcano-like workshops!!!
%To answer to these questions:
%in Radio: VLA/VLBAÉ; ALMA; in IR/optical: VLT; in HE: XMM, Chandra, Swift, Suzaku, INTEGRAL, RXTE, HESS(-2); AGILE,GLAST; RCs: AUGER; neutrinos: AMANDA, ANTARES
%\section*{Acknowledgements}
I would like to take here the opportunity to thank the organisers for
their invitation to give this review on microquasars and their role
in astroparticle physics, and also for a very nice organisation of
this workshop, in an idyllic place and atmosphere, fruitful to arise
scientific discussions and new ideas!
%of this acknowledge Elmar K\"ording, Leonardo Pelliza and Marc Rib\'o for 
%a careful rereading of the manuscript and useful discussions.
%, and sharing peaceful lunches,
%Leonardo for continuously inviting me to play football, 
%Elmar for being a nice office-mate and trying my dentist,
%and finally Nuria for taking care of me and keeping me at home
%even when I am not nice!!!

%\bibliographystyle{/Users/chaty/Library/Texmf/Bibtex/aa}
%\bibliography{/Users/chaty/Library/Texmf/Science/science}

\begin{thebibliography}{32}
\expandafter\ifx\csname natexlab\endcsname\relax\def\natexlab#1{#1}\fi

\bibitem[{{Bosch-Ramon} {et~al.}(2006){Bosch-Ramon}, {Romero}, \&
  {Paredes}}]{bosch-ramon:2006}
{Bosch-Ramon}, V., {Romero}, G.~E., \& {Paredes}, J.~M. 2006, {A\&A}, 447, 263

\bibitem[{{Bouchet} {et~al.}(1991){Bouchet}, {Mandrou}, {Roques}, {Vedrenne},
  {Cordier}, {Goldwurm}, {Lebrun}, {Paul}, {Sunyaev}, {Churazov}, {Gilfanov},
  {Pavlinsky}, {Grebenev}, {Babalyan}, {Dekhanov}, \&
  {Khavenson}}]{bouchet:1991}
{Bouchet}, L., {Mandrou}, P., {Roques}, J.-P., {et~al.} 1991, {ApJ},
  383, L45

\bibitem[{{Castro-Tirado} {et~al.}(1994){Castro-Tirado}, {Brandt}, {Lund},
  {Lapshov}, {Sunyaev}, {Shlyapnikov}, {Guziy}, \& {Pavlenko}}]{castro:1994}
{Castro-Tirado}, A.~J., {Brandt}, S., {Lund}, N., {et~al.} 1994, {ApJSS}, 92, 469

\bibitem[{{Chaty}(1998)}]{chaty:1998}
{Chaty}, S. 1998, PhD thesis, University Paris XI

\bibitem[{{Chaty} {et~al.}(2003){Chaty}, {Haswell}, {Malzac}, {Hynes},
  {Shrader}, \& {Cui}}]{chaty:2003b}
{Chaty}, S., {Haswell}, C.~A., {Malzac}, J., {et~al.} 2003, {MNRAS}, 346, 689

\bibitem[{{Chaty} {et~al.}(2001){Chaty}, {Rodr\'{\i}guez}, {Mirabel},
  {Geballe}, \& {Fuchs}}]{chaty:2001a}
{Chaty}, S., {Rodr\'{\i}guez}, L.~F., {Mirabel}, I.~F., {Geballe}, T., \&
  {Fuchs}, Y. 2001, {A\&A}, 366, 1041

\bibitem[{{Corbel} {et~al.}(2002){Corbel}, {Fender}, {Tzioumis}, {Tomsick},
  {Orosz}, {Miller}, {Wijnands}, \& {Kaaret}}]{corbel:2002}
{Corbel}, S., {Fender}, R.~P., {Tzioumis}, A.~K., {et~al.} 2002, Science, 298,
  196

\bibitem[{{Dhawan} {et~al.}(2000){Dhawan}, {Mirabel}, \&
  {Rodr\'{\i}guez}}]{dhawan:2000b}
{Dhawan}, V., {Mirabel}, I., \& {Rodr\'{\i}guez}, L. 2000, {ApJ}, 543

\bibitem[{{Distefano} {et~al.}(2002){Distefano}, {Guetta}, {Waxman}, \&
  {Levinson}}]{distefano:2002}
{Distefano}, C., {Guetta}, D., {Waxman}, E., \& {Levinson}, A. 2002,
  {ApJ}, 575, 378

\bibitem[{{Dubner} {et~al.}(1998){Dubner}, {Holdaway}, {Goss}, \&
  {Mirabel}}]{dubner:1998}
{Dubner}, G., {Holdaway}, M., {Goss}, M., \& {Mirabel}, I.~F. 1998, {AJ}, 116, 1842

\bibitem[{{Dubus}(2006)}]{dubus:2006a}
{Dubus}, G. 2006, A\&A in press

\bibitem[{Dubus \& Chaty(2006)}]{dubus:2006b}
Dubus, G. \& Chaty, S. 2006, A\&A in press

\bibitem[{{Falcke} {et~al.}(2004){Falcke}, {K{\" o}rding}, \&
  {Markoff}}]{falcke:2004}
{Falcke}, H., {K{\" o}rding}, E., \& {Markoff}, S. 2004, {A\&A},
  414, 895

\bibitem[{{Fender}(2001)}]{fender:2001a}
{Fender}, R.~P. 2001, {MNRAS}, 322, 31

\bibitem[{{Fuchs} {et~al.}(2001){Fuchs}, {Mirabel}, \& {Ogley}}]{fuchs:2001}
{Fuchs}, Y.~., {Mirabel}, I.~F.~., \& {Ogley}, R.~N. 2001, Astrop. Space SS, 276, 99

\bibitem[{{Fuchs} {et~al.}(2003){Fuchs}, {Rodriguez}, {Mirabel}, {Chaty},
  {Rib{\' o}}, {Dhawan}, {Goldoni}, {Sizun}, {Pooley}, {Zdziarski},
  {Hannikainen}, {Kretschmar}, {Cordier}, \& {Lund}}]{fuchs:2003}
{Fuchs}, Y., {Rodriguez}, J., {Mirabel}, I.~F., {et~al.} 2003, {A\&A}, 409, L35

\bibitem[{{Gallo} {et~al.}(2003){Gallo}, {Fender}, \& {Pooley}}]{gallo:2003}
{Gallo}, E., {Fender}, R.~P., \& {Pooley}, G.~G. 2003, {MNRAS}, 344, 60

\bibitem[{{Genzel} {et~al.}(2003){Genzel}, {Sch{\" o}del}, {Ott}, {Eckart},
  {Alexander}, {Lacombe}, {Rouan}, \& {Aschenbach}}]{genzel:2003}
{Genzel}, R., {Sch{\" o}del}, R., {Ott}, T., {et~al.} 2003, {Nature}, 425, 934

\bibitem[{{Heinz} \& {Sunyaev}(2002)}]{heinz:2002}
{Heinz}, S. \& {Sunyaev}, R. 2002, {A\&A}, 390, 751

\bibitem[{{Lee} {et~al.}(2002){Lee}, {Reynolds}, {Remillard}, {Schulz},
  {Blackman}, \& {Fabian}}]{lee:2002}
{Lee}, J.~C., {Reynolds}, C.~S., {Remillard}, R., {et~al.} 2002, {ApJ}, 567, 1102

\bibitem[{{Margon}(1984)}]{margon:1984}
{Margon}, B. 1984, {ARA\&A}, 22, 507

\bibitem[{{Markoff} {et~al.}(2001){Markoff}, {Falcke}, \&
  {Fender}}]{markoff:2001}
{Markoff}, S., {Falcke}, H., \& {Fender}, R. 2001, {A\&A}, 372,
  L25

\bibitem[{{Marscher} {et~al.}(2002){Marscher}, {Jorstad}, {G{\' o}mez},
  {Aller}, {Ter{\" a}sranta}, {Lister}, \& {Stirling}}]{marscher:2002}
{Marscher}, A.~P., {Jorstad}, S.~G., {G{\' o}mez}, J., {et~al.} 2002, {Nature},
  417, 625

\bibitem[{{Mirabel} {et~al.}(1998){Mirabel}, {Dhawan}, {Chaty}, {Rodriguez},
  {Marti}, {Robinson}, {Swank}, \& {Geballe}}]{mirabel:1998a}
{Mirabel}, I.~F., {Dhawan}, V., {Chaty}, S., {et~al.} 1998, {A\&A}, 330, L9

\bibitem[{{Mirabel} \& {Rodr\'{\i}guez}(1994)}]{mirabel:1994a}
{Mirabel}, I.~F. \& {Rodr\'{\i}guez}, L.~F. 1994, {Nature}, 371, 46

\bibitem[{{Mirabel} {et~al.}(1996){Mirabel}, {Rodr\'{\i}guez}, {Chaty},
  {Sauvage}, {Gerard}, {Duc}, {Castro-Tirado}, \& {Callanan}}]{mirabel:1996a}
{Mirabel}, I.~F., {Rodr\'{\i}guez}, L.~F., {Chaty}, S., {et~al.} 1996,
  {ApJ}, 472, L111

\bibitem[{{Mirabel} {et~al.}(1992){Mirabel}, {Rodr\'{\i}guez}, {Cordier},
  {Paul}, \& {Lebrun}}]{mirabel:1992b}
{Mirabel}, I.~F., {Rodr\'{\i}guez}, L.~F., {Cordier}, B., {Paul}, J., \&
  {Lebrun}, F. 1992, {Nature}, 358, 215

\bibitem[{{Rib{\' o}} {et~al.}(2004){Rib{\' o}}, {Dhawan}, \&
  {Mirabel}}]{ribo:2004}
{Rib{\' o}}, M., {Dhawan}, V., \& {Mirabel}, I.~F. 2004, in European VLBI
  Network on New Developments in VLBI Science and Technology, 111--112

\bibitem[{{Romero} \& {Orellana}(2005)}]{romero:2005}
{Romero}, G.~E. \& {Orellana}, M. 2005, {A\&A}, 439, 237

\bibitem[{{Sturner} \& {Shrader}(2005)}]{sturner:2005}
{Sturner}, S.~J. \& {Shrader}, C.~R. 2005, {ApJ}, 625, 923

\bibitem[{{T{\" u}rler} {et~al.}(2004){T{\" u}rler}, {Courvoisier}, {Chaty}, \&
  {Fuchs}}]{turler:2004}
{T{\" u}rler}, M., {Courvoisier}, T.~J.-L., {Chaty}, S., \& {Fuchs}, Y. 2004,
  {A\&A}, 415, L35

\bibitem[{{Ueda} {et~al.}(2002){Ueda}, {Yamaoka}, {S{\' a}nchez-Fern{\'
  a}ndez}, {Dhawan}, {Chaty}, \& {et al.}}]{ueda:2002}
{Ueda}, Y., {Yamaoka}, K., {S{\' a}nchez-Fern{\' a}ndez}, C., {et~al.} 2002,
  {ApJ}, 571, 918

\end{thebibliography}

% For Figures insertion uncomment the next section

%\begin{figure}
%\includegraphics{figurename}
%\caption{Your caption here}
%\label{fig01} % optional figure label, must be unique
%\end{figure}

%%%%%%%%%%%%%%%%%%%%%%%%%%%%%%%%%%%%%%%%%%%%%%%%%%%%%%%%
% End of the paper
%
\end{document}